\documentclass[10pt,a4paper]{article}

\usepackage[T1]{fontenc}
\usepackage[utf8]{inputenc}

\usepackage{graphicx}
\usepackage{booktabs}
\usepackage{listings}
\usepackage{url}
\usepackage{authblk}
\usepackage{orcidlink}
\usepackage{hyperref}
\hypersetup{
  colorlinks=true,
  linkcolor=blue,
  citecolor=blue,
  urlcolor=blue
}

\newcommand{\keywords}[1]{\par\medskip\noindent\textbf{Keywords:} #1\par}

\title{Developing Security Metrics for Automated Compliance Checking%
\thanks{This is the author-accepted manuscript of a paper published in the proceedings of IFIP SEC 2026. The final authenticated version is available from Springer Nature at \url{https://doi.org/10.1007/978-3-032-27993-4_20}.}}

\author[1]{Immanuel Kunz\orcidlink{0000-0002-4669-0030}}
\author[1]{Nico Haas\orcidlink{0000-0002-8446-8692}}
\author[1]{Angelika Schneider\orcidlink{0000-0002-8962-3276}}
\author[1]{Christian Banse\orcidlink{0000-0002-4874-0273}}

\affil[1]{Fraunhofer AISEC, Garching b.~M\"unchen, Germany}

\begin{document}

\maketitle

\begin{abstract}
To automatically check for compliance with security-related standards, prose requirements must be translated to technical properties to make them measurable. A multitude of metrics have been proposed in past literature, but they are lacking a systematic methodology for their elicitation. Without such a methodology, metrics remain inconsistent in their data models and abstraction levels, they exhibit considerable overlaps and are difficult to combine to comprehensively cover a standard's requirement.
In this paper, we develop a data model for security metrics based on a systematization of compliance requirements, and we propose a methodology for their elicitation. The method is based on an ontology that enables a systematic development of metrics with a comparable degree of abstraction. We also present our initial work on an open-source effort that demonstrates that our approach also supports a collaborative development of metrics.

\end{abstract}
\section{Introduction}
Regulatory standards as well as internal security policies require a technical representation of security requirements to make them checkable in an automatic way. Today, systems and their security state can change rapidly---a continuous, automated conformance checking is therefore necessary and, in turn, requires metrics for their technical representation.
More and more regulatory standards and certification schemes, for example the Cyber Resilience Act (CRA), European Cloud Security Scheme (EUCS), C5, etc., are being created to increase the security standard in different areas, but a common basis for their systematic representation on a technical level is missing. Moreover, these standards often have considerable overlaps, so compliance engineering could benefit from such a common basis, i.e. technical metrics. 

On the one hand, compliance standards are supposed to be generally applicable and long-lived---and thus cannot prescribe specific technologies and system-specific requirements which would be outdated soon. % Their application is also supposed to be context-specific and interpreted by humans. 
On the other hand, users of these standards, i.e., auditors and auditees, require a systematic way of making these requirements actionable and measurable.

Critically, the lack of standardization in metric development creates multiple problems: (1) metrics with inconsistent data models cannot be meaningfully compared or combined, (2) varying abstraction levels prevent systematic completeness, (3) redundant metrics emerge due to lack of common reference points, and (4) mapping metrics to multiple standards requires duplicated effort. A standardized, ontology-based approach can address these fundamental challenges while enabling collaborative development as an additional benefit.

Existing literature has put forward many security-related metrics in various fields as well as compliance engineering approaches and systems, but a common, structured approach to translating high-level requirements into technical metrics is missing.

In a previous publication, we have already proposed an ontology of cloud resources~\cite{banse2023semantic}, which we have since expanded beyond the cloud; in this paper, we build on this ontology, using it as the semantic link between requirements and metrics.
Our contributions include the following: 1) A general model of requirements as they can typically be found in certification standards, benchmarks, and compliance catalogues, 2) a metric data model with a semi-formal description of the metric's properties, and 3) an ontology-based elicitation method for metrics.
Furthermore, we therefore propose a way forward to a semi-formalized description of requirements, bridging the gap between the needs to specify requirements in an unambiguous but context-independent way. We also discuss how our approach can connect with emerging concepts like OSCAL, which aim at describing standards in a structured way.

% probably not necessary in the intro: We define security metrics not in the strict mathematical sense but as measures that in some way quantify the level of a security property.

\section{Background and Related Work}\label{sec:background-related-work}

\subsection{Terminology}
In the terminology of ISO/IEC 17065~\cite{ISO17065}, a \textit{certification scheme} defines requirements and a certification procedure. The object being certified is called \textit{product} and may, e.g., be a service, software, hardware, or a combination thereof.
In this paper, we use the word \textit{standard} in a broad sense to denote any authoritative document that prescribes requirements for a product. This includes certification schemes and benchmarks as well as control catalogues that may not involve an external audit or a resulting certificate, such as the CIS Benchmarks\footnote{\url{https://www.cisecurity.org/cis-benchmarks}}. With the method proposed in this paper, we aim at breaking down high-level certification requirements as well as lower-level benchmark requirements to similar levels.

In this paper, we use the term security metric for a measure that maps observable properties of a product to a value indicating some aspect of its security level in a reproducible way. While the mathematical term 'metric' strictly refers to distance functions satisfying specific axioms (non-negativity, identity of indiscernibles, symmetry, and triangle inequality), we adopt the broader usage common in security literature where 'metric' denotes any quantifiable measurement (cf. \cite{savola2013quality}). 

% A metric is a measure for the distance between two elements of a set and needs to fulfill non-negativity, identity of indiscernibles, symmetry, and triangle inequality to qualify as a metric. In this paper, we use the term security metric for a measure that maps observable properties of a product to a value indicating some aspect of its security level in a reproducible way.

\subsection{Security Metrics}
Work on security metrics ranges from early critical assessments of quantitative approaches to more recent proposals of quality criteria, taxonomies, surveys, and risk models.
Verendel~\cite{verendel2009quantified} provides a critical survey of 90 quantitative security models and metrics and argues that %, because of weak empirical validation and questionable assumptions, 
the validity of quantitative security approaches remains largely unclear. Subsequent work has focused on clarifying what constitutes a ``good'' metric and on structuring the space of metrics.
In contrast, our work does not attempt to validate quantitative security models, but focuses on the structured engineering of security metrics for automated compliance checking, i.e. a systematic development that Verendel calls for.
% Is from 2009 though!

Savola~\cite{savola2013quality} elicits quality criteria for security metrics based on an expert survey.
He identifies correctness, measurability, and meaningfulness as primary criteria. %, complemented by reproducibility, comparability, and objectivity.
Sherif et al.~\cite{sherif2024smart} similarly identify security metrics from the literature and assess them according to the SMART criteria (specific, measurable, actionable, relevant, timely).
Thus, these works explore what constitutes a ``good'' metric and motivate several elements of our data model (e.g., comparability and explicit purpose), but they do not provide a data model or an elicitation process.

Several authors have proposed ontologies and taxonomies to structure the space of security metrics.
Savola~\cite{savola2007towards} introduces a taxonomy that organizes metrics across business, organizational, and technical categories; he does not, however, define concrete metrics.
Longueira-Romero et al.~\cite{longueira2020quantify} systematize the landscape of security metrics for embedded systems and propose a taxonomy. % that classifies 169 metrics along assets, security dimensions, and metric properties. % So what's the difference to our work?
Pendleton et al.~\cite{pendleton2016survey} likewise survey system‑level security metrics and group them into metrics of vulnerabilities, defense power, attack severity, and security situations.
Common to these works is that their taxonomies operate at the level of the metrics themselves and aim at classifying existing metrics.
In contrast, the ontology used in this work is defined over resources and security features of the assessed systems (e.g., cloud resources and their security properties), and metrics are specified on top of this resource‑ and feature‑centric ontology by binding measurable properties of resources to standard requirements.
Rather than introducing another taxonomy of metrics, we provide a generic, ontology‑referenced data model and elicitation method that support the engineering of new metrics from standards and their machine‑readable representation for automated, continuous certification. Figure~\ref{fig:metrics-dev} in Section~\ref{sec:elicitation} depicts an excerpt from the ontology (see also \cite{banse2023semantic}).

Large-scale surveys provide further systematization of existing security metrics. Morrison et al.~\cite{morrison2018mapping} map the field of software security metrics by identifying 346 metrics from 63 papers, mapping them to software development lifecycle phases. %, including an analysis of how these metrics have been evaluated.
Pendleton et al.~\cite{pendleton2016survey}, as mentioned above, cover metrics for software, networks, economic aspects, and security effectiveness, and organize them according to their system‑level focus.
These surveys are valuable for collecting metrics, but they do not propose a systematic elicitation process or a structured, machine‑readable format.

% Other work focuses on linking metrics to cyber risk and harm. Woods and Böhme~\cite{woods2021sok} present a causal model of cyber risk and review the literature on which of the causal factors, e.g. exposure and vulnerability, can be measured, identifying a number of candidate indicators such as time‑to‑repair or system complexity.
% Their goal is to understand which indicators are suitable for quantifying cyber risk and for identifying effective security interventions.
% Our approach is complementary: we assume that desirable interventions are already specified in a standard or control catalogue and focus on their translation into technical metrics.

There are also efforts by the major cloud providers to develop specific sets of metrics, e.g. on Microsoft Azure\footnote{\url{https://github.com/Azure/azure-policy/tree/master/built-in-policies/policySetDefinitions/Regulatory\%20Compliance}}. Their policies, however, are limited to the usage in their own cloud systems. % for the CIS benchmark or other standards on AWS\footnote{\url{https://github.com/rewindio/aws-security-hub-CIS-metrics}} or 
% There is a related RFC about performance metrics~\cite{rfc6390}

Overall, existing work has explored quality criteria and taxonomies for security metrics, as well as their role in risk assessment. However, little attention has been given to the structured representation and systematic elicitation of security metrics from different standards. Our work addresses this gap by providing an elicitation method based on an ontology and a corresponding metric data model. % including an executable representation. % that together enable collaborative development and automated evaluation of security metrics in the context of certification.

% Should a metrics taxonomy be built into the ontology? Why (not)? Metrics target certain properties, so first, the resources and properties should be systematized.
    
\subsection{Continuous Certification and Measurement Approaches} 
Continuous certification approaches aim to replace point‑in‑time audits by automated evidence collection and assessment. 
They therefore rely heavily on clearly defined, automatically measurable properties that can be monitored over time.

Stephanow and Fallenbeck~\cite{stephanow2015towards} present a set of low‑level metrics for IaaS resources (e.g., CPU, memory, storage, network) and show how such metrics can be combined into higher‑level indicators for complex concepts such as scalability.
Their work demonstrates how technical measurements can support more continuous assurance of cloud infrastructures, but the metrics are defined ad hoc and without a generic data model.
In this paper, we provide a way of specifying complex metrics and we add a systematic elicitation method.

At the framework level, several EU projects have investigated continuous certification for cloud services. EU‑SEC~\cite{knoblauch2019cloud} and MEDINA~\cite{orue2021medina}, for example, developed concepts and prototypes for continuous assurance, including cloud service taxonomies and, in the case of MEDINA, ontologies for cloud resources, security features, and associated metrics. These projects provide important foundations for our work and demonstrate the feasibility of continuous certification in practice. However, the way metrics were defined and linked to standards in these projects remained largely project‑specific; they did not introduce a generic, reusable metric data model or a systematic elicitation method. Our work builds on these ontological foundations and makes this step explicit by providing a general, ontology‑referenced metric model and elicitation process that can be applied across different standards and continuous‑certification frameworks.

%\subsection{Supply Chain Security Measurement} % Not strictly necessary, but interesting related field if we have space
%\begin{itemize}
%    \item A field related to measuring security properties lies in supply chain security where the goal is to evaluate the security risk of a given dependency. These risks do not only refer to existing vulnerabilities, but also the ecosystem around a given software project. 
    % \item In this area, both academic literature and practical tooling exists. The OSSF Scorecard\footnote{\url{https://github.com/ossf/scorecard}} is a prominent example of a tool that assesses multiple project health criteria, like maintenance activity, usage of fuzzing and SAST tools, and branch protection mechanisms.
    % \item One example for a scientific approach is Gkortzis et al.~\cite{gkortzis2018vulinoss} who create a dataset of vulnerabilities in open-source projects, connected to multiple software metrics, the programming language(s), and more.
    % \item In this paper, however, we are not interested in finding meaningful security indicators, but we assume that they have already been defined in a standard.
% \end{itemize}

In summary, various works have investigated and catalogued security metrics, but little attention has been given to the structure and systematic elicitation of security metrics in the context of certification.

\section{Developing Metrics}\label{sec:developing-metrics}
In this section, we describe our method to develop metrics. We first present a model of certification requirements and build a data model for metrics. We then describe how metrics can be elicited based on an ontology. 

% \subsection{Goals} 

\subsection{A Requirement Decomposition Pattern}\label{sec:modeling-reqs}
Any single metric should target one measurable aspect of a high-level certification requirement. To develop a useful data model for metrics, we thus first create a decomposition pattern of a requirement as it is commonly found in certification objectives, benchmark controls, standard requirements, etc.

A requirement can be divided into the following parts. Note that not all of these elements are found in every requirement---more low-level benchmarks, like the CIS benchmarks for example, usually refer more directly to concrete configurations, while certification catalogues, like the CRA, rather postulate high-level requirements, since they have a broader application scope.

\begin{enumerate}
    \item \textbf{Designation:} The requirement targets an auditee which may be a cloud service provider (CSP) or a manufacturer. Examples include the following: ``The CSP shall ...'' (EUCS), ``the manufacturer shall ...'' (CRA).
    \item \textbf{Action:} The action is a verb that describes on a high level what the auditee should do, for example implement, document, revise, include, provide, etc.
    \item \textbf{Target Resource Type:} The action (2) targets a certain type of resource that is supposed to provide security properties, for example: ``... encryption mechanisms...'',  ``... completed by all internal and external employees ...'' (C5). Note that the target resource type is sometimes ambiguous. A requirement about employee trainings, for instance, can be interpreted as targeted at employees or at the execution of the training. In such cases, it is helpful to focus on the aspect that shall be measured later; in this case, it is the completion of the training. Note also that the resource type is not always explicitly mentioned and it is assumed that all applicable resources are meant.
    \item \textbf{Means:} The action (2) is specified by more concrete means that should be applied. Consider the following examples: ``... that corresponds to the state-of-the-art ...'' (EUCS), ``... including the following aspects ...'' (C5).
    \item \textbf{Scope}: Often, the scope of the requirement is further specified as in the following example: ``... that are used to deploy the cloud service in the production environment ...'' (C5).% , ``... for the transmission of CSC data over public networks ...'' (EUCS). % ``... unless otherwise agreed between manufacturer and business user ...'' (CRA).
    \item \textbf{Purpose:} To make the requirement better understandable for the reader, they often state the purpose of the requirement, for example: ``... to limit attack surfaces ...'' (CRA), ``... in order to protect the confidentiality ...'' (EUCS), ``... in order to avoid administrative errors ...'' (C5).
\end{enumerate}

We call a requirement decomposed into the above elements a \textit{systematized requirement}. Before we can derive metrics from this decomposition, we first analyze its mapping to our ontological knowledge base. This mapping enforces a common abstraction level and prevents redundancy in metrics development.
% To arrive at meaningful metrics, we handle the six parts described above as follows:
\begin{itemize}
    \item The \textbf{designation} (1) is not relevant for the metric, since it does not carry information for the measurement.
    \item The \textbf{action} (2) is not directly reflected in a metric, but it may give an indication about the target resource type.
    \item The \textbf{target resource type} (3) is defined in our ontology. It includes taxonomy of resources, including code, documentation, (cloud) resources, etc., which should reflect the requirement's target resource type and its properties. % This is referenced in the metric's description field.
    \item The \textbf{means} (4) are primarily reflected in the ontology's security feature taxonomy. This taxonomy is ordered by security goals, like confidentiality, integrity, etc., and defines mechanisms that target these goals, like at-rest encryption, RBAC, and rate limiting. These concepts are also detailed with more properties, for example if the encryption is enforced, which algorithm it uses, etc. 
    \item A metric should then use these ontology concepts to bind together a resource type and the necessary means. The \textbf{scope} (5) may also be reflected in the ontology concepts or they need to be used in the application scope of the metric, i.e., in the compliance engineering system that is used.
    \item Finally, the \textbf{purpose} (6) is not directly relevant for the metric but should be included in its general description to clarify its intention.
    % \textit{target resource type} shall have \textit{property (defined in scope)}
\end{itemize}

\subsection{Modeling Metrics}\label{sec:modeling-metrics}
% TODO: add identifiers to the requirements and pick them up later; human-readable -> semi-structured description; identifiable->ID and version; configurability -> separate meta and config data; practical categorization -> category; context/purpose -> comments
As seen in Section~\ref{sec:background-related-work}, metrics should aim for various qualities, including meaningfulness, measurability, specificity, and correctness \cite{savola2013quality}. The data model for security metrics should enable and support these goals. Furthermore, consider the following practical requirements to be important:

% \item These data model-related requirements can only provide the framework for defining a useful metric. Considering a concrete instance of a metric, it should pursue the following goals: Savola's correctness, measurability, and meaningfulness. A good metric should be conceptually specific, quantitatively measurable, practically attainable, consistently measured without subjective criteria, and time-dependent % N. Pham, L. Baud, P. Bellot, and M. Riguidel, “A Near Real-Time System for Security Assurance Assessment,” in Proceedings of the 2008 The Third International Conference on Internet Monitoring and Protection, Washington, DC, USA, 2008, pp. 152–160.

% \subsubsection{Readability} 
% \subsubsection{Allow Complexity}

\begin{enumerate} 
    \item From the pattern described in the previous section, we can derive that a metric should unambiguously refer to the ontology concepts to specify what is being measured. This way, a collaborative development with a common knowledge base can be enabled.
    \item Considering the multitude of stakeholders that are involved in compliance systems, metrics should both be human-readable, for instance for governance roles without technical background, and automatically processable.
    \item The language that is used to describe a metric must also allow to combine target properties via logical operators to allow for complex expressions, for instance when certain assumptions need to be expressed.
    \item A purpose and context description should be included to enable users to quickly grasp how the metric should be applied.
    \item For practicality, the metrics should furthermore provide a clear identifier, including a version specifier. This identification is necessary because at a later point in time auditors may want to spot check certain assessment results and need the exact metric version that had been used to make the assessment.
    \item Metrics include long-lived metadata, like identifiers and the ontology-based description, as well as relatively short-lived configuration data, like the target value. These should clearly be marked and divided, so they can independently be handled if necessary.
\end{enumerate}

Based on the requirements established above, we now present a comprehensive description language and data model for metrics. The design of this framework required consideration of multiple representation formats, each offering distinct trade-offs between human readability and machine processability.

The spectrum of available description languages ranges from free-text formats, which maximize human readability at the expense of deterministic machine interpretation, to formal logic languages such as Prolog, which provide unambiguous machine-readable specifications but sacrifice accessibility for non-technical stakeholders. Semi-structured formats occupy an intermediate position, offering a balance between these competing concerns.

We adopt a hybrid approach that addresses the needs of diverse stakeholders. The representation employs a semi-structured format that maintains human readability while mandating the integration of concepts from our ontological basis. This design decision ensures a common description language accessible to governance professionals, domain experts, and other non-technical personnel who must understand metric definitions. By enforcing references to ontological concepts, the semi-structured format promotes consistency while avoiding the opacity of purely formal representations. Table~\ref{tab:datamodel} presents the data model.

\begin{table}
  \centering
  \begin{tabular}{p{0.18\textwidth}p{0.8\textwidth}} 
    \toprule
    \textbf{Metadata} & \\
    \toprule
     ID & A unique identifier for the metric which is also human-readable \\
     Description & A semi-formalized description of the metric which unambiguously refers to ontology concepts and the configuration data (see below) \\
     Category & A thematic category this metric belongs to \\
     Version & Number that unambiguously refers to a certain version of this metric \\
     Comments & Further comments, e.g., about the purpose of the metric and possibly the context in which it is meaningful \\
    \toprule
    \multicolumn{2}{p{\textwidth}}{\textbf{Configuration Data}} \\
    \toprule
    Property 1 & \\
    - Operator & The operator to be used for the assessment \\
    - Target Value & The desired target value \\
    \hline
    Property X & \\
    - Operator & The operator to be used for the assessment \\
    - Target Value & The desired target value \\
    \bottomrule
  \end{tabular}
  \caption[]{The data model of a security metric.}\label{tab:datamodel}
\end{table}

Recognizing that semi-structured descriptions inevitably contain ambiguities that preclude fully automated processing, we complement each metric with a deterministic, machine-readable specification implemented in the Rego policy language\footnote{See \url{https://www.openpolicyagent.org/docs/policy-language}}. The Rego implementation serves as an executable translation of the metric definition, enabling automated evaluation, validation, and enforcement. 
Importantly, our architecture does not impose exclusive dependency on Rego; the framework permits incorporation of additional implementation languages as needed to accommodate organizational preferences or technical constraints.

This dual-representation strategy offers several advantages. First, it provides format-appropriate interfaces tailored to different stakeholder roles: governance professionals interact with structured natural language, while engineers work with executable code implementations. Second, it enables independent evolution of human-facing documentation and machine-executable logic. Third, it facilitates validation and verification by enabling cross-checking between human-readable specifications and executable implementations, reducing the risk of misinterpretation or implementation errors. The representation also allows for more complex expressions where a target value may, for example, be dependent on a limitation. Note that from the experience of aligning this data model across different stakeholders and projects (see Section~\ref{sec:repo}), it has undergone multiple improvement iterations already.
% TODO: How do we ensure that the Rego version correctly implements the metric?

%\begin{itemize}
%    \item 
%    % \item TODO Rego example?
%\end{itemize}

%\begin{itemize}
%    \item Metadata
%        \begin{itemize}
%            \item \textbf{ID:} EncryptionForSensitiveResourcesEnabled
%            \item \textbf{Description:} This rule assesses whether a [Storage] has  [AtRestEncryption] [p1:enabled] correctly configured, if the respective [Storage] has [p2:label] assigned.
%            \item \textbf{Category:} StorageEncryption
%            \item \textbf{Version:} 1.0
%            \item \textbf{Comments:} At rest encryption is a standard practice to keep sensitive data confidential when it is stored permanently.
%        \end{itemize}
%    \item Configuration Data
%    \begin{itemize}
%        \item p1:
%            \begin{itemize}
%                \item \textbf{Operator:} ==
%                \item \textbf{Target Value:} True
%            \end{itemize}
%        \item p2:
%            \begin{itemize}
%                \item \textbf{Operator:} ==
%                \item \textbf{Target Value:} "sensitive"
%            \end{itemize}
%    \end{itemize}
%    \item Note that from the experience of aligning this data model across different stakeholders and projects (see Section~\ref{sec:repo}), it has undergone multiple improvement iterations already.
%\end{itemize}

\subsection{Elicitation Steps}\label{sec:elicitation}
There are two starting points: bottom-up and top-down, i.e., either develop metrics based on what can practically be measured in a given system or derive metrics from high-level standards. In both approaches, we systematically add identified properties to the ontology, creating a comprehensive model of the possible target of evaluation. This process ensures that no redundant metrics are introduced, since one needs to find or introduce the required property in the ontology first.

% Identify technical property in standard requirement, in the ontology, in a use-case or in other practical observations
Note that alternative approaches exist. For example, one could start from the ontology itself and identify gaps in the resource descriptions and security features by comparing them to existing descriptions of the respective resources. For instance, one may compare an existing cloud resource description with the respective cloud resource description by a major cloud provider\footnote{Using, e.g., the Azure REST reference which specifies many resource properties.}.

Regardless of the chosen approach, metrics should clearly address the requirement's target resource type and the properties that are relevant for the specified scope:

\begin{enumerate}
    \item We start the elicitation by dividing the requirement into the elements identified earlier, i.e., Designation, Action, Target Resource Type, Means, Scope, and Purpose. This ensures all relevant aspects are explicitly captured.
    \item Model the requirement using the template to create a formal representation linking components to ontological concepts.
    \item Identify the specific resource and security properties that need to be measured or verified for compliance (technical or organizational).
    \item Map the identified properties to existing concepts in the ontology. If needed, extend the ontology while maintaining consistency.
    \item Formalize the metric by defining its evaluation logic and implement it using Rego code. The metric specifies how to collect evidence and determine compliance.
\end{enumerate}
    
\begin{figure} 
  \includegraphics[width=12.2cm]{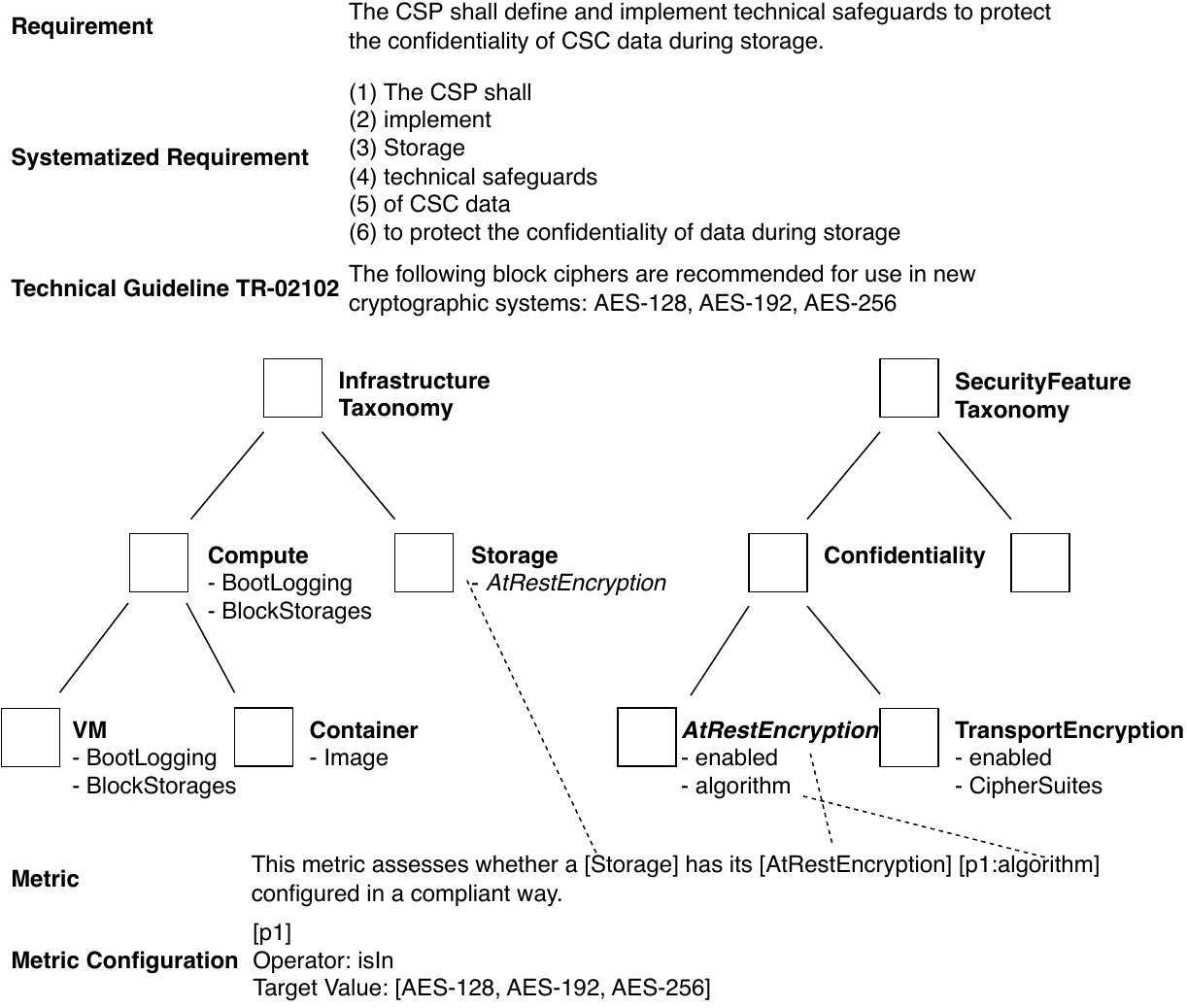}
  \caption{The metrics development approach: First, requirements are broken down into the six elements described in Section \ref{sec:modeling-reqs} (systematized requirement). Possibly, one may use technical guidelines published by authorities to help make the target resource type and the means more precise. The figure then shows two related excerpts of the ontology: On the left, the infrastructure taxonomy which structures resource types and their properties, and on the right, the security feature taxonomy which structures security properties and mechanisms. The infrastructure taxonomy may specify that certain resources offer security features, for example a Storage offers AtRestEncryption---whose properties (like the algorithm) are specified in the taxonomy on the right. 
  % This decoupling of resources and their security features allows to flexibly refer to these properties. 
  Finally, we can derive a metric from the systematization: We define a metric that refers to the Storage ontology term and its AtRestEncryption feature. The metric configuration then offers a way to define the concrete target values to be considered compliant.} \label{fig:metrics-dev}
\end{figure}

The ontology-based elicitation method systematically prevents the problems identified in Section 1. By requiring each metric to reference ontology concepts, the approach ensures: (1) no duplicate metrics are created for the same property, (2) all metrics operate at comparable abstraction levels defined by the ontology structure, (3) gaps in requirement coverage become visible through ontology analysis (identifying ontology properties which are not referenced in any metric), and (4) metrics from different sources can be integrated through shared ontological vocabulary. Figure~\ref{fig:metrics-dev} illustrates an example of this process.
% Example: We want to measure compliance with the certification requirement from the EUCS OPS-13.1B ``The CSP shall store all log data in an integrity-protected and aggregated form that allow its evaluation.'' (top-down)
% The ontology\footnote{Referring to the version from 21.10.2025, see http://...} already includes logging, a storage target, and Immutability as an Integrity security feature.

Metrics can also become more complex, e.g. when a condition is included in the requirement. In this case, the elicitation steps may need to be repeated for the condition, and the resulting metrics need to be connected via a logical AND or OR. % TODO: e.g. if logging enabled, then target has integrity protection enabled
Note that we do not claim that every requirement can be translated comprehensively to technical metrics. We rather aim at finding meaningful targets that can reduce burden on manual verification. We still suggest to define metrics also for properties that cannot (yet) be measured automatically. In this case, evidence can be introduced manually and be managed alongside automatically collected evidence. Metrics may thus be technical or organizational.

\subsection{Example}
We exemplify our method using a requirement from the C5 standard (BSI C5 CRY-02) which reads: "The Cloud Service Provider has established procedures and technical measures for strong encryption and authentication for the transmission of data of cloud customers over public networks." Following the pattern from Section~\ref{sec:modeling-reqs}, we decompose this requirement into its elements:

\begin{enumerate}
    \item \textbf{Designation}: The Cloud Service Provider
    \item \textbf{Action}: has established
    \item \textbf{Target Resource Type}: Resources that offer TransportEncryption
    \item \textbf{Means}: technical measures for strong encryption
    \item \textbf{Scope}: for the transmission of data of cloud customers over public networks
    \item \textbf{Purpose}: --
\end{enumerate}

It is noteworthy that we have made the resource type explicit by adding "Resources that offer TransportEncryption" which was only implicitly mentioned in the original requirement text. This demonstrates a key benefit of the ontology-based elicitation method: it enforces a common semantic basis by requiring explicit mapping to ontology concepts. In our ontology, resources are defined with a property "offers TransportEncryption" when they can be configured to provide transport-level encryption. This explicit mapping creates greater reproducibility and consistency across different metric implementations.

Based on this systematization, we can now derive a concrete metric. The metric is described as follows: \textit{This rule assesses whether a [Resource] has its [TransportEncryption] [p1:protocol] correctly configured.}
The brackets indicate references to ontology concepts, ensuring that the metric unambiguously specifies what is being measured. The parameter [p1:protocol] represents a configurable value that will be evaluated against specified target values, e.g. \textit{TLS}.

In the Rego implementation of this metric, the application scope can be appropriately limited to resources that possess the respective transport encryption property. This ensures that the metric is only applied to relevant resources. Note also that the limitation regarding cloud customer data over public networks should not be embedded within the metric definition itself. Instead, this limitation should be handled by the compliance system that applies the metric. This separation of concerns allows the metric to remain general and reusable while enabling context-specific application through the compliance checking framework. % This approach ensures that the same metric can be applied flexibly across different deployment scenarios while maintaining the specific scoping requirements of individual standards or organizational policies.

\subsection{A Collaborative Security Metrics Repository}\label{sec:repo}
The ontology-based elicitation of metrics enables a collaborative development of metrics. We have created an open-source repository for this purpose. We have founded the Cybersecurity Certification Hub (CCH) which governs the repository, available on GitHub\footnote{\url{https://github.com/Cybersecurity-Certification-Hub/security-metrics}}. Currently, multiple EU-funded projects are contributing to the repository\footnote{COBALT (\url{https://horizon-cobalt.eu}), EMERALD (\url{https://www.emerald-he.eu/}), and CONFIRMATE (\url{https://confirmate-project.eu/}).}
The CCH is working on extending the repository with metrics related to, e.g., the EUCS\footnote{\url{https://www.enisa.europa.eu/publications/eucs-cloud-service-scheme}}, AIC4\footnote{\url{https://www.bsi.bund.de/EN/Themen/Unternehmen-und-Organisationen/Informationen-und-Empfehlungen/Kuenstliche-Intelligenz/AIC4/aic4\_node.html}}, CRA\footnote{\url{https://digital-strategy.ec.europa.eu/en/policies/cyber-resilience-act}}, and other standards and is open for contributions from other projects and organizations. Existing categories of metrics include, e.g., AI, network, and application security.

\section{Discussion}\label{sec:discussion}
% This work presents a systematic approach to developing security metrics for automated compliance checking, addressing the need for structured, collaborative metric development in the context of continuous certification. The following subsections discuss practical implications, limitations, and future directions.

\subsection{Practical Implications}
The ontology-based approach facilitates collaborative development of metrics across stakeholders and projects and advances existing efforts to systematize metrics development~\cite{longueira2020quantify,pendleton2016survey,savola2007towards}. The metrics repository is currently integrated into the Confirmate platform\footnote{See \url{https://github.com/confirmate}}, enabling automated processing of metrics. Both top-down (requirement-driven) and bottom-up (measurement-driven) approaches have proven valuable in practice. While we favor the top-down approach to ensure comprehensive coverage of requirements, bottom-up metrics that measure meaningful properties can typically be mapped to existing controls as well.

The proposed metrics enable automated compliance checking, which is increasingly required by modern standards. For instance, the EUCS mandates that "The CSP shall perform at regular intervals [...] audits by subject matter experts to check the compliance of their internal security control system" (CO03), while the C5 catalog explicitly acknowledges automation potentials in many requirements. Also, the recent draft of C5 2025 acknowledges this and enhances machine-readability of its requirements.

%\begin{itemize} % TODO review
  % \item Does the metric now fulfill the requirements? How well does the method allow collaborative development? % TODO
  % \item Decoupling the metric definition and the code-level implementation (in Rego) strikes a trade-off between keeping the metric understandable for different stakeholders and complete machine-readability.
%\end{itemize}

\subsection{Limitations}
Several limitations should be considered. First, automated compliance checking is limited by Goodhart's law: metrics may cease to effectively measure security when used as key performance indicators, as auditees may optimize toward fulfilling metrics rather than implementing comprehensive security measures. Second, security cannot be fully captured by any finite set of metrics---they remain approximations of complex security properties. 
Yet, while prose requirements interpreted by human auditors may better capture contextual nuances, manual audits face different limitations: auditors may focus disproportionately on easily verifiable indicators while overlooking less tangible security aspects.

Our approach, however, offers flexibility in abstraction levels. Metrics can range from low-level configuration checks to high-level aggregators evaluating complex testing procedures. This flexibility allows tailoring metric granularity to specific assessment contexts. 
Nevertheless, automatically checking multiple metrics cannot fully encompass the scope of prose-written requirements intended for human interpretation. This represents an inherent trade-off between automation efficiency and assessment comprehensiveness. Evaluating this trade-off empirically requires comparing the coverage achieved by metric-based assessments against traditional audit outcomes.

\subsection{Complementing OSCAL}
The ontology-based metric description complements emerging standardization efforts such as OSCAL (Open Security Controls Assessment Language)\footnote{\url{https://pages.nist.gov/OSCAL}}. OSCAL provides structured representations of security catalogs, controls and assessment procedures and has gained adoption with several US standards, such as NIST SP 800-53. Also European institutions consider the usage of OSCAL, for example the German BSI provides its basic security catalogue ("IT-Grundschutz") in the OSCAL format\footnote{\url{https://github.com/BSI-Bund/Stand-der-Technik-Bibliothek}}.

Central to OSCAL is the \textit{Control Layer}, including its \textit{Catalog Model}, which enables machine-readable descriptions of security catalogs and governance frameworks. It provides structured textual descriptions of controls, splitting control prose into statements. These statements suit the atomic requirements needed in our elicitation step (see~Section~\ref{sec:elicitation}). Statements can be parametrized when needed, e.g. for defining review intervals. These parameters could also be used to link to metrics and specify their target values. % Each control has a unique identifier for establishing well-defined mappings from security controls to implementing or assessing metrics.

OSCAL also defines an \textit{Assessment Layer} with models for \textit{Assessment Plans} and \textit{Assessment Results}. This layer complements our metric model through several elements. It contains activities, each with a defined method and detailed steps, clearly linked to related controls. Unlike our approach, it focuses on human-readable task descriptions versus machine-executable policies. However, activities and steps can include user-defined properties with key, value, class and namespace, allowing well-defined links to the metric repository via metric IDs. The Assessment Plan model specifies components and assets for assessment. While focusing on concrete asset instantiation, it allows specifying \textit{asset types} that could link into the ontology using well-defined terms (e.g., [VirtualMachine]). Integrating our metric model into OSCAL would thus enable machine-actionable compliance frameworks spanning from high-level control definitions to concrete technical measurements.

\subsection{Towards a Semi-Formalized Certification Scheme}
The ontology-based description framework developed here offers a path toward semi-formalized certification schemes. Current requirements like ``The CSP shall define and implement procedures and technical safeguards to protect the confidentiality of CSC data during storage'' (EUCS CKM-03) could be structured as: ``The CSP shall document [encryption] mechanisms in [document] resources'' and ``The CSP shall configure [AtRestEncryption] as [enabled] on [Storage] resources containing CSC data.''

Such ontology-based requirement formulations would maintain human readability and contextual flexibility while clearly specifying target resources and properties. This systematization would simplify metric derivation and enable governance bodies to centrally update referenced measurement properties as the state of the art evolves. Furthermore, structured resource descriptions in ontologies enable automated tooling for conformance checking processes~\cite{banse2024owl2proto}.

\section{Conclusions}
Systematic standardization of security metric development is essential to address fundamental challenges in compliance engineering, like inconsistent data models, varying abstraction levels, metric redundancy, and mapping requirements across multiple standards. While prior work has catalogued numerous security metrics, a structured methodology for their standardized development has been missing. This paper addresses this gap through an ontology-based approach that ensures consistency and comprehensiveness, prevents duplication, and enables cross-standard applicability. Collaborative development emerges as a benefit of this standardization framework which we have started in an open-source repository. Furthermore, we have proposed a way towards semi-formalized standards, using the ontological basis.

% TODO This work presents a systematic approach to developing security metrics for automated compliance checking, facilitating several benefits: proliferating metrics development, reducing redundancy, enabling cross-standard efficiency, ensuring systematic completeness, supporting evolution and governance, and bridging stakeholder perspectives.

% In this paper, we have developed a model of certification requirements, a data model for security metrics, and an elicitation method for metrics. The data model strikes a trade-off between a human-readable format and a machine-readable, automatically processable one.

In future work, we will develop more metrics, map them to standards, and move towards an empirical evaluation in a practical environment. Given the structured templates, data models, ontology, and methodology presented in this paper, future work will also explore LLM-assisted translation of prose requirements to formal metrics.
% TODO: also, investigate how to design comprehensive metric suites (e.g. based on Wagner).

\section*{Acknowledgments}
This work was partly funded by COBALT (Grant agreement ID: 101119602), EMERALD (Grant agreement ID: 101120688), and CONFIRMATE (project number: 101190193).

\section*{Competing Interests}
The authors have no competing interests to declare that are relevant to the content of this article.

%
% ---- Bibliography ----
%
% References will be sorted and formatted in the correct style.
%
\bibliographystyle{splncs04}
\bibliography{bib}
\end{document}